# A new algebraic technique for polynomial-time computing the number modulo 2 of Hamiltonian decompositions and similar partitions of a graph's edge set.

**Greg Cohen**
**Nizhny Novgorod State University**
**greg.cohen.math@gmail.com**

In Graph Theory a number of results were devoted to studying the computational complexity of the number modulo 2 of a graph's edge set decompositions of various kinds, first of all including its Hamiltonian decompositions, as well as the number modulo 2 of, say, Hamiltonian cycles/paths etc. While the problems of finding a Hamiltonian decomposition and Hamiltonian cycle are NP-complete, counting these objects modulo 2 in polynomial time is yet possible for certain types of regular undirected graphs. Some of the most known examples are the theorems about the existence of an even number of Hamiltonian decompositions in a 4-regular graph and an even number of such decompositions where two given edges $e$ and $g$ belong to different cycles (Thomason, 1978), as well as an even number of Hamiltonian cycles passing through any given edge in a regular odd-degreed graph (Smith's theorem). The present article introduces a new algebraic technique which generalizes the notion of counting modulo 2 via applying fields of Characteristic 2 and determinants and, for instance, allows to receive a polynomial-time formula for the number modulo 2 of a 4-regular bipartite graph's Hamiltonian decompositions such that a given edge and a given path of length 2 belong to different Hamiltonian cycles – hence refining/extending (in a computational sense) Thomason's result for bipartite graphs. This technique also provides a polynomial-time calculation of the number modulo 2 of a graph's edge set decompositions into simple cycles each containing at least one element of a given set of its edges what is a similar kind of extension of Thomason's theorem as well. Additionally, it gives a polynomial-time algorithm for the number modulo 2 of a graph's edge set decompositions into simple paths such that each vertex is an end of exactly one path, as well as the number modulo 2 of such decompositions into odd simple paths and into even simple paths.

In the paper all the considered undirected graphs are assumed to be loopless (unless mentioned otherwise), while they may contain multiple edges.

**Theorem (in any field of Characteristic 2):**

*Let G=(V,E) be an undirected graph all whose vertices are of even degrees; U be the adjacency matrix of the arc-weighted digraph whose vertex set is the 2|E| directed edges of G such that the entries* $u_{(i,j),(j,k)} = r_{i,k}^{[j]}$ *, where* $R^{[j]}$ *is a symmetric unitary* $\deg(j)\times\deg(j)$*-matrix with a zero diagonal, and all the other entries are equal to zero; w be a |E|-vector of weights of edges of G,* $\vec{w}$ *be the 2|E|-vector of weights of directed edges of G equal to the weights of the corresponding undirected ones.*

*Then*

$$\det\left(U^{\bullet 2} + Diag(\vec{w})\right) = \Big( \sum_{D\in Eulerdec(G)} \Big( \prod_{((i,j),(j,k))\in D} r_{i,k}^{[j]}\Big) \prod_{C\in D} \big(1 + \prod_{(i,j)\in C} w_{(i,j)}\big)\Big)^2$$

*where Eulerdec(G) is the set of decompositions of E into cycles with no repeated edges,*

$U^{\bullet 2}$ *denotes* $\{u_{ij}^2\}_{m\times m}$.

***Proof:***

first of all, for an $m\times m$-matrix $A$ and an $m$-vector $d$ let's define the cycle polynomial $cycle(A,d) =^{def}= \sum_{\pi\in S_m}(\prod_{i=1}^{m} a_{i\pi_i})\prod_{C\in Cyc(\pi)}(1+\prod_{i\in C} d_i)$ where $Cyc(\pi)$ is the set of directed cycles of the permutation $\pi = \begin{pmatrix} 1 & \dots & m \\ \pi_1 & \dots & \pi_m \end{pmatrix}$. It's easy to see that

$$cycle(A,d) = \sum_{I\subseteq M}\det(A^{(I,I)})(\prod_{k\in M\setminus I} d_k)\det(A^{(M\setminus I,M\setminus I)})$$

where $M=\{1,...,m\}$ and for $X\subseteq M$ $A^{(X,X)}$ denotes the sub-matrix of $A$ whose sets of rows and columns both are $X$.

If $A$ is unitary, i.e. $AA^T = \mathrm{I}_m$, then $\det(A^{(I,I)}) = \det(A)\det((A^{-1})^{(M\setminus I,M\setminus I)}) = \det((A^T)^{(M\setminus I,M\setminus I)}) =$ $= \det(A^{(M\setminus I,M\setminus I)})$ (due to the equalities det($A$)=1 and $A^{-1} = A^T$) and hence $cycle(A,d) =$ $= \sum_{I\subseteq M}\det^2(A^{(I,I)})\prod_{k\in M\setminus I} d_k = \sum_{I\subseteq M}\det((A^{\bullet 2})^{(I,I)})\prod_{k\in M\setminus I} d_k = \det\left(A^{\bullet 2} + Diag(d)\right)$ where $A^{\bullet 2}$ denotes $\{a_{ij}^2\}_{m\times m}$ and the applied equality $\det^2((A)^{(I,I)}) = \det((A^{\bullet 2})^{(I,I)})$ is due do Characteristic 2. By the construction, $U$ is unitary as there exists a pair of permutations of its rows and columns reducing it to $Diag(\begin{pmatrix} R^{[1]} \\ \dots \\ R^{[|V|]} \end{pmatrix})$. Hence $cycle(U,\vec{w}) = \det\left(U^{\bullet 2} + Diag(\vec{w})\right)$.

If we define $opposite\left((i,j)\right) =^{def}= (j,i)$ then we can vertex-wise subject any system of the digraph's directed cycles to this mapping hence receiving a partition of the set of these systems into pairs and "singles" (equal to their opposite ones). Due to the symmetry the equalities $u_{(i,j),(j,k)} = u_{(k,j),(j,i)}$, $\vec{w}_{(i,j)} = \vec{w}_{(j,i)} = w_{(i,j)}$ generate, a system of directed cycles and its opposite one (if these are two different systems) have the same entries $u_{(i,j),(j,k)}$-s and $\vec{w}_{(i,j)}$-s. Hence we can take into account (because of Characteristic 2) only those systems which are mapped onto themselves ("singles"), namely the systems where each directed cycle is paired with its opposite one. Because $u_{(i,j),(j,i)} = 0$ for all $(i,j)\in E$, those systems consist of directed cycles which can't contain both $(i,j)$ and $(j,i)$, hence any directed cycle of theirs paired with its opposite one forms an undirected edge-unrepeated cycle of the original graph $G$ (as any undirected edge-unrepeated cycle of $G$ has precisely two directions of going around). Eventually we get $cycle(U,\vec{w}) = \sum_{D\in Eulerdec(G)}(\prod_{((i,j),(j,k))\in D} u_{(i,j),(j,k)})^2\prod_{C\in D}(1+\prod_{(i,j)\in C} w_{(i,j)})^2$ where *Eulerdec(G)* is the set of decompositions of $E$ into undirected cycles with no repeated edges (or the Eulerian cycles of its edge-subgraphs), what completes the proof of Theorem.

---

**Definition**: for an undirected graph *G=(V,E)* each of whose vertices *j* of degree deg(*j*) is assigned a symmetric $\deg(j)\times\deg(j)$-matrix $R^{[j]}$ and each of whose edges $(i,j)$ is assigned a weight $w_{(i,j)}$ over a field of Characteristic 2, let's define the cycle-decomposition polynomial by

$$cycledec(G,R^{[1]},...,R^{[|V|]},w) =^{def}= \sum_{D\in Eulerdec(G)} (\prod_{((i,j),(j,k))\in D} r_{i,k}^{[j]})\prod_{C\in D}(1+\prod_{(i,j)\in C} w_{(i,j)})$$

**Denotation:** for a natural number *m* let's denote by $\overline{1}_m$ the *m*-vector all whose entries are 1.

To be symmetric, unitary, and zero-diagonaled, $R^{[j]}$ is constructed as $I_{\deg(j)}+(P^{[j]})^T P^{[j]}$ where $P^{[j]}$ is a $q\times\deg(j)$-matrix such that $P^{[j]}(P^{[j]})^T = 0_{q\times q}$, $(P^{[j]})^T\overline{1}_q = \overline{1}_{\deg(j)}$. Particularly, if each $P^{[j]} = \overline{1}_{\deg(j)}^T$ then we receive the primitive case of $cycledec(G,w) =^{def}= \sum_{D\in Dec(G)}\prod_{C\in D}(1+\prod_{(i,j)\in C} w_{(i,j)})$ where *Dec(G)* is the set of decompositions of *E* into simple cycles (because in such a case all the passage coefficients are equal to 1, while any undirected graph which isn't a simple cycle has an even number of Eulerian cycles).

Generally, if we call a vertex with $P^{[j]} = \overline{1}_{\deg(j)}^T$ simple-cycled then we can claim that

$$cycledec(G,R^{[1]},...,R^{[|V|]},w) = \sum_{D\in Eulerdec(G,V_{simple})} (\prod_{((i,j),(j,k))\in D} r_{i,k}^{[j]})\prod_{C\in D}(1+\prod_{(i,j)\in C} w_{(i,j)})$$

where $V_{simple}$ is the set of simple-cycled vertices , *Eulerdec(G,* $V_{simple}$*)* is the set of decompositions of *E* into cycles with no repeated edges and no repeated simple-cycled vertices (or the Eulerian cycles of its edge-subgraphs with simple-cycled vertices of degree 2 or 0). Let's call $r_{i,k}^{[j]}$ the passage coefficient between *i* and *k* through *j*. Due to the passage coefficients, we can solve in polynomial time (modulo randomization, i.e. the class RP), for instance, such problems as the decomposability of an undirected graph (edge-weighted over a field of Characteristic 2) into edge-unrepeated cycles such that the products of their edge-weights are not 1 -- it follows directly from the cycle-decomposition polynomial's definition and the linear independence of all the possible products of passage coefficients which can be generated by a cycle-decomposition, in the generic case of $P^{[j]}$, $P^{[j]}(P^{[j]})^T = 0_{q\times q}$.

**Lemma 1**: *if G=(V,E) is a 2h-regular graph,* $w_{(i,j)} = 1+\lambda_{(i,j)}\varepsilon$ *then* $coeff_{\varepsilon^h} cycledec(G,w) = \sum_{D\in Hamdec(G)}\prod_{C\in D}\sum_{(i,j)\in C}\lambda_{(i,j)}$ *where Hamdec(G) is the set of Hamiltonian decompositions of G. Particularly, in the case of all the lambdas equal to 1, we receive the number modulo 2 of Hamiltonian decompositions for graphs with an odd number of vertices.*

**Lemma 2**: *if G=(V,E) is a graph, F is a subset of E,* $\overline{1}_{E\setminus F}$ *is the |E|-vector whose entries are 0 when indexed by elements of F and 1 otherwise then* $cycledec(G,\overline{1}_{E\setminus F})$ *is the number modulo 2 of decompositions of E into simple cycles containing at least one element of F.*

Let's call an edge ***supporting*** if its weight isn't 1 and let's define the set of supporting edges of a cycle *C* as the ***supporting set*** of *C*. We'll also say that a cycle *C* is ***supported*** by its supporting set if the product of its weights isn't 1 and is ***not supported*** otherwise.

**Lemma 2.1:** *if G=(V,E) is a 4-regular graph of order n, e, $g_1, g_2$ are its pair-wise distinct edges,* $w_{(i,j)} = 1\ for\,(i,j) \notin \{e, g_1, g_2\},\ w_e = \alpha^{-1},\ w_{g_1} = w_{g_2} = \alpha \neq 1$ *then* $\frac{\alpha}{(1+\alpha)^3} cycledec(G,w)$ *is the number modulo 2 of decompositions of E into a simple cycle of length n-1 or n containing e and either two simple cycles each containing one of the edges $g_1, g_2$ and vertex-disjoint in all their vertices except possibly one or a Hamiltonian cycle containing both $g_1, g_2$.*

***Proof :***

the proof of this lemma is based on Thomason's theorem about the existence of an even number of Hamiltonian decompositions of a quartic multi-graph with at least two vertices where two given edges belong to different cycles. The theorem implies that any graph has an even number of its edge set's decompositions into two simple cycles possessing more than one common vertex such that two given edges belong to different cycles -- because any connected graph whose edge set is decomposable into two simple cycles such that two given edges are in different cycles is homeomorphic to a quartic multi-graph (with possible loops) where the two given edges of the original graph would transform into two edges of the new multi-graph and such pairs of cycles of the original graph would one-to-one correspond to Hamiltonian pairs of the new multi-graph such that those two new edges are in different cycles. Let's also notice that a disconnected graph whose edge set is decomposable into two simple cycles is a pair of disjoint simple cycles. Hence, given a simple cycle *C* of *G* containing *e* and not containing $g_1$ or/and $g_2$, the graph $G_{\setminus C} = (V, E \setminus C)$ (i.e. the rest of *G*'s edge set) can have an odd number of decompositions into two simple cycles such that $g_1, g_2$ are in different cycles only in case if it's homeomorphic to a pair of loops (possibly connected), hence being a pair of simple cycles with at most one common vertex such that $g_1, g_2$ are in different cycles. It completes dealing with the case where edges *e*, $g_1, g_2$ belong to three pair-wise distinct cycles with each cycle containing exactly one element of the set $\{e, g_1, g_2\}$.

Therefore there remains, due to the equalities $w_e w_{g_1} = w_e w_{g_2} = 1$, just the option of *e* belonging to one cycle and both $g_1, g_2$ to another one (as a quartic graph's edge set is decomposable into not fewer than two simple cycles, while *e*, $g_1, g_2$ are the only supporting edges and, by the definition of *cycledec(G,w)*, we can consider only decompositions of *E* into cycles having at least one of them and supported by their supporting sets – but cycles with the supporting sets $\{e, g_1\}$ and $\{e, g_2\}$ are not supported by them). By the latter conclusion, the whole proof of Lemma is completed too since in both cases (of two and three cycles) we receive the cycle-decomposition polynomial's summands $(1+\alpha)^2(1+\alpha^{-1}) = \frac{(1+\alpha)^3}{\alpha}$ corresponding to the cycle-decompositions of *G* mentioned in Lemma.

---

If *G* is bipartite and hence can't have a cycle of length *n-1* (where *n* is its order) then we immediately receive

**Corollary 2.1.1:** *if G=(V,E) is a 4-regular bipartite graph, e, $g_1, g_2$ are its pair-wise distinct edges,* $w_{(i,j)} = 1\ for\,(i,j) \notin \{e, g_1, g_2\},\ w_e = \alpha^{-1},\ w_{g_1} = w_{g_2} = \alpha \neq 1$ *then* $\frac{\alpha}{(1+\alpha)^3} cycledec(G,w)$ *is the number modulo 2 of decompositions of E into a Hamiltonian cycle containing e and either* *two simple vertex-disjoint cycles each containing one of the edges* $g_1, g_2$ *or a second Hamiltonian cycle containing both* $g_1, g_2$.

And at last, in case if $g_1$, $g_2$ are adjacent and hence can't belong to two different vertex-disjoint cycles, we receive the following ***refinement*** of Thomason's theorem:

**Corollary 2.1.2:** *if G=(V,E) is a 4-regular bipartite graph, e,* $g_1$, $g_2$ *are its pair-wise distinct edges, {* $g_1$, $g_2$ *} is a path of length 2,* $w_{(i,j)} = 1 \; for \, (i,j) \notin \{e, g_1, g_2\}$, $w_e = \alpha^{-1}$, $w_{g_1} = w_{g_2} = \alpha \neq 1$ *then* $\frac{\alpha}{(1+\alpha)^3} cycledec(G,w)$ *is the number modulo 2 of G's Hamiltonian pairs where e and the path {* $g_1$, $g_2$ *} belong to different cycles.*

**Definition:** let's call a cycle ***q-simple*** in the vertex *j* if it goes through it at most *q* times.

**Lemma 2.2:** *if G=(V,E) is a graph,* $F_1, F_2$ *are two disjoint sets of its edges,*

$$w_e = \begin{cases} 1, e \notin F_1 \cup F_2 \\ \alpha^{-1}, e \in F_1 \\ \alpha, e \in F_2 \end{cases}$$

*where* $\alpha$ *is a formal variable then* $coeff_{\alpha^{|F_2|}} cycledec(G,w)$ *is the number modulo 2 of decompositions of E into simple cycles each containing at least one element of* $F_1 \cup F_2$ *but not containing edges from them both.*

***Proof:***
to prove this lemma, we just need to notice that, due to the cycle-decomposition polynomial's definition, under the conditions given in Lemma $cycledec(G,w)$ is a polynomial in $\alpha$ of degree $|F_1| + |F_2|$ divided by $\alpha^{|F_1|}$.

**Definition:**
for a graph *G=(V,E)* and two disjoint sets of its edges $F_1, F_2$ let's denote by $cycledec_{F_1,F_2}(G)$ the number modulo 2 of decompositions of *E* into simple cycles each containing at least one element of $F_1 \cup F_2$ but not containing edges from them both.

**Corollary 2.2.1:**
*let G=(V,E) be a graph,* $\{e_1, g_1\}, ..., \{e_m, g_m\}$ *be pair-wise disjoint pairs of its edges.*
*Then* $\sum_{F \in \{e_1,g_1\} \times ... \times \{e_m,g_m\}, e_1 \in F} cycledec_{F, \{e_1,g_1\} \cup ... \cup \{e_m,g_m\} \setminus F}(G)$ *is the number modulo 2 of decompositions of E into cycles each containing at least one element of* $\{e_1, g_1\} \cup ... \cup \{e_m, g_m\}$ *and not containing any pairs* $\{e_1, g_1\}, ..., \{e_m, g_m\}$ *such that the corresponding multi-graph whose vertices are these cycles and whose edges are formed by the pairs* $\{e_1, g_1\}, ..., \{e_m, g_m\}$ *is connected and bipartite.*

**Definition:** let A be a $q \times q$-matrix, then $ham(A) =^{def}= \sum_{\pi \in S_q^{[ham]}} \prod_{i \in \pi} a_{i,\pi_i}$, where $S_q^{[ham]}$ is the set of Hamiltonian $q$-permutations.

From the definition of the cycle polynomial (given in the proof of Theorem), it follows that $ham(A) = cycle(A, \begin{pmatrix} 0 \\ \overline{1}_{q-1} \end{pmatrix}) = \sum_{I \subseteq \{1,\ldots,q\}, 1 \in I} \det(A^{(I,I)}) \det(A^{(\{1,\ldots q\} \setminus I, \{1,\ldots,q\} \setminus I)})$ and hence $ham(A)$ is polynomial-time computable for any unitary $A$.

**Lemma 3:**

*Let for j=1,…,|V|* $rank(P^{[j]}) = q_j$, $R^{[j]} = I_{\deg(j)} + (P^{[j]})^T P^{[j]}$.

*Then* $cycledec(G, R^{[1]}, \ldots, R^{[|V|]}, w) = \sum_{D \in Eulerdec(G, q_1, \ldots, q_{|V|})} \left( \prod_{((i,j),(j,k)) \in D} r_{i,k}^{[j]} \right) \prod_{C \in D} \left(1 + \prod_{(i,j) \in C} w_{(i,j)}\right)$ *where* $Eulerdec(G, q_1, \ldots, q_{|V|})$ *is the set of decompositions of E into edge-unrepeated cycles which are* $q_j$*-simple in the vertex j, j=1,…,|V| (or (*$q_1, \ldots, q_{|V|}$*)-simple cycle decompositions).*

***Proof:***

By the definition, $cycledec(G, R^{[1]}, \ldots, R^{[|V|]}, w) = \sum_{D \in Eulerdec(G)} \left( \prod_{((i,j),(j,k)) \in D} r_{i,k}^{[j]} \right) \prod_{C \in D} \left(1 + \prod_{(i,j) \in C} w_{(i,j)}\right)$ .

Let's take a cycle-decomposition from *Eulerdec(G)* with a cycle *C* which goes $q$ times through the vertex *j*. Let $q > q_j$ and the neighbors of *j* in the cycle *C* be $i_1, k_1, \ldots, i_q, k_q$ (which are pair-wise distinct due to having no repeated edges), while the other patterns of the cycle *C* (the "interpaths" between those neighbors) be between $k_1$ and $i_1$ ,…, $k_q$ and $i_q$. We can denote them by $Path(k_1, i_1)$,…, $Path(k_q, i_q)$ correspondingly (let's denote this set of paths by *Interpath(C,j)* and call it a *j*-Interpath). Then there are *(q-1)!* edge-unrepeated cycles consisting of the edges *C* consists of (i.e. the set *E(C)*) and containing *Interpath(C,j)*. Let's call these cycles *j*-equivalent to *C*. By the *j*-equivalence relation of cycles (which partitions the set of cycles of *G*), *Eulerdec(G)* is also partitioned into sets of *(j,Interpath(C,j))*-equivalent cycle-decompositions which differ only in the cycle with the same *j*-Interpath and only in the vertex *j* .

For the sets $K = \{k_1, \ldots, k_q\}$, $I = \{i_1, \ldots, i_q\}$ and $T \subseteq \{1, \ldots, q\}$ let's denote by $K_T, I_T$ the subsets of $K, I$ correspondingly consisting of their elements with sub-indexes from *T,* by $K_{\setminus T}, I_{\setminus T}$ the sets $K \setminus K_T$, $I \setminus I_T$ correspondingly. Let's define the passage coefficient of a cycle through a vertex as the product of its related passage coefficients through the vertex. Then the sum of the *(j,Interpath(C,j))*-equivalent cycles' passage coefficients through $j$ is

$$\sum_{T, 1 \in T} ham((R^{[j]})^{(K_T \cup I_{\setminus T}, K_{\setminus T} \cup I_T)}) = \sqrt{\sum_{T, 1 \in T} \det((R^{[j]})^{(K_T \cup I_T, K_T \cup I_T)}) \det((R^{[j]})^{(K_{\setminus T} \cup I_{\setminus T}, K_{\setminus T} \cup I_{\setminus T})})}$$

The last equality is due to the symmetry of $R^{[j]}$. The expression in the left side doesn't depend on the diagonal entries of $R^{[j]}$, hence the right side can be re-written as

$$\sqrt{\sum_{T, 1 \in T} \det((R^{[j]} + I_{2q})^{(K_T \cup I_T, K_T \cup I_T)}) \det((R^{[j]} + I_{2q})^{(K_{\setminus T} \cup I_{\setminus T}, K_{\setminus T} \cup I_{\setminus T})})}$$

By the construction, $rank(R^{[j]} + I_q) = q_j < q$. Thus, since at least one of the two cardinalities $|I_T \cup K_T|$, $|I_{\setminus T} \cup K_{\setminus T}|$ exceeds $q_j$, at least one of the two determinants $\det((R^{[j]})^{(K_T \cup I_T, K_T \cup I_T)})$, $\det((R^{[j]})^{(K_{\setminus T} \cup I_{\setminus T}, K_{\setminus T} \cup I_{\setminus T})})$ is equal to zero. Therefore, given a cycle-decomposition $D \in$ *Eulerdec(G),* upon supposing the pair (*j,Interpath(C,j)*) to be its lexicographical minimum among all such pairs (*x,Interpath(Y,x)*) with the condition $\deg_Y(x) > q_x$ (where *Y* is a cycle of *D* having the vertex *x*, $\deg_Y(x)$ is the number of times *Y* goes through *x*), it completes the proof of Lemma because if *D* isn't ($q_1, \ldots, q_{|V|}$)-simple then it has such a unique lexicographical minimum (*j,Interpath(C,j)*) and the corresponding *(j,Interpath(C,j))*-equivalence class doesn't intersect any other equivalence class. Hence all the ($q_1,..., q_{|V|}$)-nonsimple summands of the cycle-decomposition polynomial's expansion are partitioned into sets whose sums are zeroes.

**Lemma 4**: *Let G=(V,E) be a graph,* $\lambda$ *be an* $|E|$*-vector,* $rank(P^{[j]}) = q_j$,
$R^{[j]} = \mathrm{I}_{\deg(j)} + (P^{[j]})^T P^{[j]}$ *for j=1,…,*$|V|$, $coeff_{\varepsilon^h} cycledec(G, R^{[1]},..., R^{[|V|]}, \overline{1}_{|E|} + \varepsilon\lambda) \neq 0$.
*Then there exists a decomposition of E into at most h (*$q_1,..., q_{|V|}$*)-simple cycles.*

**Definition:**
let's define a graph's ***normal cycle-path-decomposition*** as a partition of its edge set into cycles and paths with no repeated edges such that each vertex is an end of exactly one path.

**Definition:**
given a graph *G=(V,E),* let's define the ***doubling*** of *G* as two copies of *G* with each pair of corresponding vertices connected by an edge, i.e. the graph *Double(G)=(*$V' \cup V''$*,Double(E))* where $|V'| = |V''| = |V|$, both sub-graphs induced by $V', V''$ are *G* (under the vertex maps $i' \to i, i'' \to i$ correspondingly) and the edges between $V', V''$ are only $(1',1''),…, (|V|',|V|'')$. Let's call the vertex map $i' \to i'', i'' \to i'$ its ***intercopy-mapping***. Let's also consider a directed edge *(i,j)* as the ***returning*** passage *((i,j),(j,i))* and say that a path which ends in the vertex *j* and goes through the undirected edge *(i,j)* has this passage (which we'll also call an ***endedge***).

**Lemma 5:**
$$\sqrt{cycledec(Double(G), \hat{R}^{[1']}, \hat{R}^{[1'']},..., \hat{R}^{[|V|']}, \hat{R}^{[|V|'']}, \hat{w})} =$$
$$= \sum_{D \in Nortmaleulerdec(G)} \Big( \prod_{((i,j),(j,k)) \in D} r_{i,k}^{[j]} \Big) \Big( \prod_{C \in D} (1 + \prod_{(i,j) \in C} w_{(i,j)}) \Big) \prod_{P \in D} (1 + \sqrt{w_{end_1(P)} w_{end_2(P)}} \prod_{(i,j) \in P} w_{(i,j)})$$
*where* $\hat{w}_{(i',j')} = \hat{w}_{(i'',j'')} = w_{(i,j)}$, $\hat{w}_{(i',i'')} = w_i$, $\hat{r}^{[j']}_{(i',k')} = \hat{r}^{[j'']}_{(i'',k'')} = r^{[j]}_{(i,k)}, \hat{r}^{[j']}_{(i',j'')} = \hat{r}^{[j'']}_{(i'',j')} = r^{[j]}_{(i,j)}$,
$Normaleulerdec(G)$ *is the set of normal cycle-path-decompositions of G, C denotes a cycle, P denotes a path whose end-vertices are* $end_1(P), end_2(P)$.

***Proof:***
the proof of this lemma is based on the symmetry of *Double(G)* and resembles the proof of the present article's main theorem. By the intercopy-mapping, the set of cycle-decompositions of *Double(G)* is partitioned into "pairs" and "singles" where each "pair" consists of its two cycle-decompositions having the same product of their passage-coefficients while each "single" is two copies of a normal cycle-path-decomposition of *G* with the ends of corresponding paths connected by edges $(i', i'')$.

Now, for an undirected graph whose vertices, edges, paths of length 2 (passages) and directed edges (endedges) are weighted over a field of Characteristic 2, let's define the cycle-path-decomposition polynomial by $cyclepathdec(G, R^{[1]},..., R^{[|V|]}, w) =^{def} =$
$$=^{def} = \sum_{D \in Normaleulerdec(G)} \Big( \prod_{((i,j),(j,k)) \in D} r_{i,k}^{[j]} \Big) \Big( \prod_{C \in D} (1 + \prod_{(i,j) \in C} w_{(i,j)}) \Big) \prod_{P \in D} (1 + \sqrt{w_{end_1(P)} w_{end_2(P)}} \prod_{(i,j) \in P} w_{(i,j)})$$

All the above theory for the cycle-decomposition polynomial can be transferred to the cycle-path-decomposition polynomial as well. First of all, if each passage-coefficient matrix $R^{[j]}$ (whose rows and columns correspond to the vertex $j$ and its neighbors) is unitary, symmetric and zero-diagonaled then it's polynomial-time computable because in such a case the corresponding matrices $\hat{R}^{[j'']}$, $\hat{R}^{[j'']}$ are unitary, symmetric and zero-diagonaled too as the index maps $i' \to i, i'' \to i$ transform them into $R^{[j]}$ (by re-indexing their rows and columns). Let's call all *G*'s passages except its endedges *proper* and denote the proper passage-coefficient matrices by $R^{[j,proper]}$. Then we get the following extension of Lemma 4 (upon defining a path *q*-simple in a vertex in the same manner that we did for a cycle when considering only proper passages through the vertex as going through it):

**Lemma 4.1 :**
*Let G=(V,E) be a graph,* $w_e = 1+\varepsilon\lambda_e$ *for* $e \in E$, $w_i = 1+\varepsilon^2\lambda_i$ *for* $i \in V$,
$rank(R^{[j,proper]} + I_{\deg(j)}) = q_j$ *for j=1,...,|V|,* $coeff_{\varepsilon^h} cyclepathdec(G, R^{[1]},...,R^{[|V|]}, w) \neq 0$.
*Then there exists a normal decomposition of E into at most h (* $q_1,..., q_{|V|}$ *)-simple cycles and paths.*

By analogy with the cycle-decomposition polynomial, if all the passage-coefficients (including those of endeges) are equal to one then we get the primitive case of

$$cyclepathdec(G,w) =^{def}= \sum_{D \in Normaldec(G)} \left(\prod_{C \in D}(1+\prod_{(i,j)\in C} w_{(i,j)})\right)\prod_{P \in D}(1+\sqrt{w_{end_1(P)}w_{end_2(P)}}\prod_{(i,j)\in P} w_{(i,j)})$$

where *Normaldec(G)* is the set of normal simple cycle-path-decompositions of *G*, i.e. decompositions of its edge set into simple cycles and paths such that each vertex is an end of exactly one path. In such a case we immediately receive

**Lemma 5.1:**
$cyclepathdec(G,w) = \sqrt{cycledec(Double(G),\hat{w})}$ where $\hat{w}_{(i',j')} = \hat{w}_{(i'',j'')} = w_{(i,j)}$, $\hat{w}_{(i',i'')} = w_i$

Particularly, if *G* is a graph of an even order *n*, $w_{(i,j)} = 1+\varepsilon$ and $w_i = 1+\alpha\varepsilon$ for all *i,j* then $coeff_{\varepsilon^{n/2}} cyclepathdec(G,w)$ is the number modulo 2 of its edge set's normal decompositions into simple paths each having an odd number of edges for $\alpha = 0$ and an even number of edges for $\alpha = 1$.

At last, if we slightly modify the definition of a normal cycle-path-decomposition into the definition of an ***odd-normal*** cycle-path-decomposition defined as a partition of a graph's edge set into cycles and paths with no repeated edges such that each odd-degreed vertex is an end of exactly one path and the definition of the doubling of a graph into the definition of the ***odd-doubling*** defined as its two copies with each pair of corresponding odd-degreed vertices connected by an edge then we can modify all the above theory for the cycle-path-decomposition polynomial into an analogous one for the odd cycle-path-decomposition polynomial $oddcyclepathdec(G, R^{[1]},..., R^{[|V|]}, w) =^{def}=$

$$=^{def}= \sum_{D \in Oddnormaleulerdec(G)} \left(\prod_{((i,j),(j,k))\in D} r^{[j]}_{i,k}\right)\left(\prod_{C \in D}(1+\prod_{(i,j)\in C} w_{(i,j)})\right)\prod_{P \in D}(1+\sqrt{w_{end_1(P)}w_{end_2(P)}}\prod_{(i,j)\in P} w_{(i,j)})$$

where *Oddnormaleulerdec(G)* is the set of odd-normal cycle-path-decompositions of *G*.